\documentclass[a4paper,11pt]{article}

\usepackage[margin=1in]{geometry}
\usepackage{authblk}      
\usepackage{hyperref}     
\usepackage[margin=1in]{geometry} 
\usepackage{amsmath}              
\usepackage{graphicx}             
\usepackage{hyperref}             
\usepackage{mathtools}

\title{\textbf{Differentiating a HEP Analysis Pipeline within the Scikit-HEP Software Ecosystem}}

\author[1]{Mohamed Aly\thanks{\href{mailto:mohamed.aly@cern.ch}{mohamed.aly@cern.ch}}}
\author[1]{Lino Gerlach\thanks{\href{mailto:lino.oscar.gerlach@cern.ch}{lino.oscar.gerlach@cern.ch}}}

\affil[1]{\large\textit{Princeton University, Princeton, NJ 08544, USA}}

\date{}

\begin{document}

\maketitle

\begin{abstract}
A first differentiable analysis pipeline is presented for an example high-energy physics (HEP) use case with publicly available collision data from the Compact Muon Solenoid detector at the Large Hadron Collider. The pipeline combines tools from the Scikit-HEP ecosystem with \texttt{JAX}. The study is based on an existing search for a hypothetical particle, the $Z^{\prime}$ boson, and uses a realistic, yet simplified, statistical model. The gradient-based optimization techniques employed in this work can advance HEP workflows by enabling end-to-end tuning of analysis parameters, improving both computational scalability and overall sensitivity. The challenges of adopting such techniques in HEP workflows are highlighted, along with practical mitigation to those challenges. This framework results in a significant improvement in expected statistical significance compared to a baseline analysis by fine-tuning $\mathcal{O}(10^3)$ parameters in the pipeline. Perspectives on future applications and recommendations for broader engagement with differentiable techniques in the field are also outlined.
\end{abstract}

\section{Introduction}

The Standard Model (SM) of particle physics is the most precise framework describing the fundamental constituents of matter and their interactions \cite{SM_review}. Experiments at the Large Hadron Collider (LHC) \cite{LHC_design}, notably the Compact Muon Solenoid (CMS) detector \cite{CMS_detector}, have confirmed many of its predictions with high accuracy~\cite{Higgs_discovery_ATLAS,Higgs_discovery_CMS}. The SM consists of matter particles (quarks and leptons), their anti-matter counterparts, force carriers (gluons, photons, $W$ and $Z$ bosons), and the Higgs boson. In CMS, these particles are not observed directly but inferred from the signatures they leave in the detector: quarks and gluons produce concentrated sprays of energy called \textit{jets}, whereas electrically charged leptons (e.g. electrons) create distinct paths with small energy deposits known as \textit{tracks}. Photons create localized showers in the calorimeters, while neutrinos, which are uncharged leptons, leave no direct signal and their presence is inferred through an apparent imbalance in the momentum measurement, called the missing transverse momentum $E_T^{\text{miss}}$. Of particular relevance to this work are the \textit{top} quark, the heaviest known fundamental particle, which decays almost immediately into a \textit{bottom} quark and a $W$ boson, and the \textit{muon}, a charged lepton that provides a clean experimental signature. Despite its success, the SM is believed to be incomplete, as it does not explain multiple observed phenomena~\cite{DM_review,Baryogenesis_review}. This motivates ongoing searches for signs of physics Beyond the Standard Model (BSM), including potentially undiscovered particles. 

A typical high-energy physics (HEP) analysis follows a complex, data-intensive, multi-step pipeline. It begins with raw detector signals, proceeds through the reconstruction of identifiable particle signatures such as tracks, jets, photons, and missing transverse momentum, and ends with a statistical test of a physics hypothesis. Machine learning (ML) methods are now employed at nearly every stage of this pipeline, from generative models for simulation, through detector calibrations and the reconstruction of tracks and jets, to the classification of entire collision events based on derived quantities that summarize event characteristics~\cite{novel-ml-lhc}. Traditionally, however, each ML component is trained as a standalone task, optimized only for its target objective. This leaves potential performance gains untapped, since the components are not optimized with respect to the ultimate goal of a HEP analysis, such as maximizing the sensitivity of a new-particle search.

In other data-intensive domains, such as computer vision, there has been a shift from pipelines of multiple, purpose-built algorithms to large, end-to-end deep learning models that learn directly from low-level inputs. For instance, modern video recognition systems process raw pixel sequences with deep neural networks rather than relying on separate, frame-by-frame feature extraction steps. This raises an important question for HEP: should analyses move toward unified models that take detector-level data as input and directly optimize for the final physics objective, such as the sensitivity of a search?

Such an end-to-end, black-box model is unlikely to be desirable for scientific research, since physicists must be able to validate and interpret each stage of an analysis. A more powerful and transparent alternative is to preserve the traditional modular structure of a physics workflow, that is calibrating detectors, reconstructing signatures in subdetectors, combining them into particles, and performing statistical inference, while formulating the entire pipeline as a single differentiable program. In this way, the workflow can be optimized holistically for the final physics objective, while its structure embeds domain knowledge and maintains interpretability. Each intermediate step remains accessible for validation, combining the strengths of a modular analysis with the advantages of gradient-based optimization~\cite{diffhep}. This hybrid approach is enabled by the paradigm of \textbf{differentiable programming}. Deep learning is one example of this paradigm: its training relies on backpropagation, a form of reverse-mode automatic differentiation \footnote{Reverse-mode automatic differentiation is a method for calculating derivatives in a way that is both exact and efficient. Unlike finite-difference approaches, automatic differentiation applies the chain rule directly to the sequence of operations that define the computation. In reverse mode, the derivatives of all intermediate quantities are propagated backwards from the output to the inputs, so that the gradient of a single scalar with respect to many parameters can be obtained in one pass.}, that efficiently computes gradients of a loss function with respect to millions of parameters. Differentiable programming generalizes this idea beyond ML models to arbitrary programs, enabling end-to-end optimizable workflows in which gradients can flow through the entire chain of computations. The \textbf{neos} framework has demonstrated this concept in a toy HEP analysis~\cite{neos}.

In the presented work~\footnote{This document is a summary a talk given at the Fifth MODE Workshop on Differentiable Programming for Experiment Design. The slides can be found  \href{https://indico.cern.ch/event/1481852/contributions/6464860/}{here}.}, these ideas are extended to a realistic setting by implementing a differentiable version of a search for the $Z^{\prime}$ boson, a hypothetical massive particle predicted by many BSM theories, which decays into a top and anti-top quark pair ($t\bar{t}$)~\cite{ttbar}. This case study builds on the analysis setup developed during the CMS Open Data Workshop in 2024~\cite{cms-opendata-workshop-2024}, using publicly available CMS collision data (CMS open data). With tools from the Scikit-HEP ecosystem, a fully differentiable analysis pipeline is constructed and its parameters optimized to maximize discovery significance. This work sits at the intersection of differentiable programming and physics-informed machine learning, pointing towards more powerful yet interpretable analysis strategies in high-energy physics.

\section{Differentiable analysis pipeline}

This study focuses on the final stage of the HEP workflow: applying event selections, constructing histograms, and performing statistical inference. The starting point of that stage is reconstructed events, from which event selection criteria are applied and histograms of key observables are built. These histograms define the inputs to a binned statistical model, from which a profile likelihood~\footnote{The profile likelihood is used to evaluate a specific parameter by maximizing the likelihood over all other parameters for each possible value of that parameter.} is constructed. Hypothesis tests based on this likelihood then yield statistical results such as exclusion limits or discovery significances. The statistical result is considered the objective of a physics analysis.
The contribution here is to express this entire chain, from event selection through statistical inference, as a differentiable program, such that gradients can be propagated from the physics objective back through the analysis. This enables the gradient-based optimization of analysis parameters directly for the objectives of a physics analysis.

\subsection{The \texorpdfstring{$Z' \to t\bar{t}$}{Z' to ttbar} analysis: a case study}
\label{sec:zprime-analysis}

The chosen case study for differentiable HEP workflows is a simplified version of a search for the $Z'$ boson. As previously mentioned, the $Z'$ boson decays into a $t\bar{t}$ pair. In turn, each top quark decays into a $W$ boson and a bottom quark. The focus of this work is placed on the single-lepton final state, where one $W$ boson decays into a muon and a neutrino, while the other decays into quarks.

Collision events that can mimic the $Z'$ signal are known as background events. The two main background sources for the signal are the production of $t\bar{t}$ pairs and the production of a $W$ boson in association with jets ($W$+jets) from SM interactions. In general, the discovery potential of an analysis is optimized by applying selection criteria (cuts) on the collision events, with thresholds chosen to reduce backgrounds while retaining as much signal as possible. The fraction of signal events passing the cuts is known as the signal efficiency. The simplified version considered here, based on the analysis presented at the CMS Open Data Workshop~\cite{cms-opendata-workshop-2024}, uses the following cuts: exactly one muon, $E_T^{\text{miss}}$ greater than 50 GeV as a proxy for the undetected neutrino, at least one $b$-tagged jet \footnote{\emph{$b$-tagging} refers to algorithms that identify jets originating from bottom quarks.}, and exactly one large-radius jet \footnote{A \emph{large-radius jet} is a jet reconstructed with a larger energy clustering parameter so that it can capture all the collimated energy deposits from the decay products of a heavy particle, such as the top quark.} consistent with a top quark decaying into lighter quarks. A transverse energy $H_T$, defined as the scalar sum of the muon momentum and $E_T^{\text{miss}}$, greater than 150 GeV is also required. The final discriminating variable is the reconstructed invariant mass of the $t\bar{t}$ system ($m_{t\bar{t}}$), where a $Z'$ signal would appear as a localized resonance, or “bump”, on top of a smoothly falling background distribution.

\subsection{Differentiable primitives and optimizable parameters}

A significant challenge in creating a differentiable HEP analysis is that many standard operations are inherently non-differentiable. For instance, applying a cut corresponds to a step function, which has zero gradient almost everywhere and an undefined gradient at the threshold. To address this, such discrete operations are replaced with smooth, differentiable \textit{relaxations}. As such, a hard cut can be approximated by a sigmoid function, with a \textit{temperature} parameter controlling the sharpness of the transition, so that the gradient of the selection efficiency with respect to the cut value becomes well defined. Similarly, histograms are non-differentiable since a small change in the value of a bin entry may cause it to migrate into another bin. This is overcome by using a binned kernel density estimate (bKDE), where each entry is represented by a Gaussian kernel and the bin content is obtained by integrating the sum of these kernels over the bin width. The kernel bandwidth then controls the smoothness of the resulting histogram.

In the analysis considered in this work, the gradients are computed with respect to a set of analysis parameters $\boldsymbol{\alpha}$ that govern how events are selected and modeled. These include the thresholds of selection cuts as well as parameters that define differentiable primitives, such as the bandwidth of the bKDEs. More generally, $\boldsymbol{\alpha}$ may also include the weights of ML models integrated into the pipeline, allowing classifiers to be re-optimized jointly with the rest of the analysis. By treating all of these quantities as differentiable parameters, the entire pipeline can be optimized directly for the final physics analysis objective.

\subsection{The statistical model and physics objective}
\label{sec:implicit_diff}
The final step of the analysis is a statistical test based on a binned profile likelihood constructed from the $m_{t\bar t}$ histogram,
\[
\mathcal{L}(\mathbf{n}\mid \boldsymbol{\alpha},\boldsymbol{\theta})
= \prod_{i\in \mathcal{B}} \mathrm{Pois}\!\left(n_i \,\middle|\, \nu_i(\boldsymbol{\alpha},\boldsymbol{\theta})\right)
\times \prod_{j\in \mathcal{J}} c_j(\theta_j).
\]
Here, $\mathbf{n} = \{n_i\}$ are the observed counts in each histogram bin $i$, and 
$\nu_i(\boldsymbol{\alpha},\boldsymbol{\theta})$ are the predicted counts in that bin, which depend on the  analysis parameters $\boldsymbol{\alpha}$ and additional fit parameters $\boldsymbol{\theta}$. 
The function $\mathrm{Pois}(n_i \mid \nu_i)$ denotes the Poisson probability of observing $n_i$ events when $\nu_i$ are expected. The vector $\boldsymbol{\theta}$ includes both the parameters of interest (e.g.\ the signal strength) and additional parameters that account for experimental and theoretical uncertainties. These uncertainty parameters, often called \emph{nuisance parameters}, allow the model to adapt to such effects. 
A subset of them, indexed by $\mathcal{J}$, are associated with external calibration or control measurements. Their influence is introduced through constraint terms $c_j(\theta_j)$, which are typically Gaussian or log-normal 
functions encoding how well those parameters are known from this auxiliary information.
To computationally simplify the optimization, the negative log-likelihood $\ell$ is utilized:
\[
\ell(\boldsymbol{\alpha},\boldsymbol{\theta}) \;\coloneqq\; -\log \mathcal{L}(\mathbf{n}\mid \boldsymbol{\alpha},\boldsymbol{\theta}).
\]
For a fixed choice of $\boldsymbol{\alpha}$, the fitted values of $\boldsymbol{\theta}$, denoted as $\hat{\boldsymbol{\theta}}$, are obtained by minimizing $\ell$:
\[
\hat{\boldsymbol{\theta}}(\boldsymbol{\alpha}) \;=\; \arg\min_{\boldsymbol{\theta}}\; \ell(\boldsymbol{\alpha},\boldsymbol{\theta})\,.
\]
This corresponds to solving the condition
\[
\nabla_{\boldsymbol{\theta}}\,\ell(\boldsymbol{\alpha},\boldsymbol{\theta})
\;\Big|\;_{\boldsymbol{\theta}=\hat{\boldsymbol{\theta}}(\boldsymbol{\alpha})}
= \mathbf{0}.
\]
The physics objective of interest, $S(\boldsymbol{\alpha})$, depends on $\boldsymbol{\alpha}$ both explicitly and implicitly through the $\hat{\boldsymbol{\theta}}$, i.e. $ S\big(\boldsymbol{\alpha}, \hat{\boldsymbol{\theta}}(\boldsymbol{\alpha})\big).$

During a gradient-based optimization of $\boldsymbol{\alpha}$, the fitted parameters are typically extracted by performing the likelihood fit at every step of the optimization, which is computationally expensive. Instead, \textit{fixed-point differentiation}, based on the implicit function theorem, is applied in the present work to compute the dependence of $\hat{\boldsymbol{\theta}}$ on $\boldsymbol{\alpha}$ directly. 
The theorem gives
\[
\frac{d\,\hat{\boldsymbol{\theta}}}{d\boldsymbol{\alpha}}
\;=\;
-\,\Big[\frac{\partial^2 \ell}{\partial \boldsymbol{\theta}\,\partial \boldsymbol{\theta}^\top}\Big]^{-1}
\frac{\partial^2 \ell}{\partial \boldsymbol{\alpha}\,\partial \boldsymbol{\theta}^\top}\,,
\]
which can be directly inserted in the expression for the total derivative of $S$ with respect to $\boldsymbol{\alpha}$. Consequently, the gradient of $S(\boldsymbol{\alpha})$ can be obtained without repeatedly solving the likelihood minimization problem inside the optimization loop. This makes large-scale, end-to-end HEP analysis optimization computationally tractable. In this work, $S(\boldsymbol{\alpha})$ is taken to be the $p$-value from a profile likelihood ratio test, i.e.\ the probability, under the background-only hypothesis, of obtaining data at least as signal-like as observed. The test is evaluated using the asymptotic approximations shown in~\cite{Cowan:2010js}.

\section{Implementation}

\subsection{The setup}
The differentiable analysis presented in this work is implemented using tools from the Scikit-HEP ecosystem and \texttt{JAX}. 
The \texttt{awkward-array} library is used during the data preparation stage for handling the jagged data structures common in HEP, and \texttt{vector} is used calculations involving Lorentz vectors. 
To compute observables and formulate differentiable selection cuts, a pure \texttt{JAX} implementation is employed~\footnote{The source code is publicly available at \href{https://github.com/ligerlac/z-prime-ttbar-gradients}{https://github.com/ligerlac/z-prime-ttbar-gradients}.}.
Gradient descent on the discovery $p$-value is performed using the \texttt{OptaxSolver} from \texttt{jaxopt}, with tailored learning rates for different parameter groups and bounds imposed on parameter values to ensure they remain within physically meaningful ranges.

The optimization is performed simultaneously over both analysis and fit parameters. 
The analysis parameters $\boldsymbol{\alpha}$ include the thresholds of key selection cuts: 
the $E_T^{\text{miss}}$ cut, the $b$-tagging score threshold used to count $b$-jets, and the $H_T$ cut. 
The fit parameters $\boldsymbol{\theta}$ are the signal strength $\mu$ and the normalization factor $\kappa_{t\bar{t}}$ for the SM $t\bar{t}$ background. 
The statistical model is implemented entirely in \texttt{JAX} to preserve differentiability, with the profile likelihood fits and hypothesis testing performed using the \texttt{relaxed} library via implicit differentiation.

\begin{figure}
    \centering
    \includegraphics[width=0.49\textwidth]{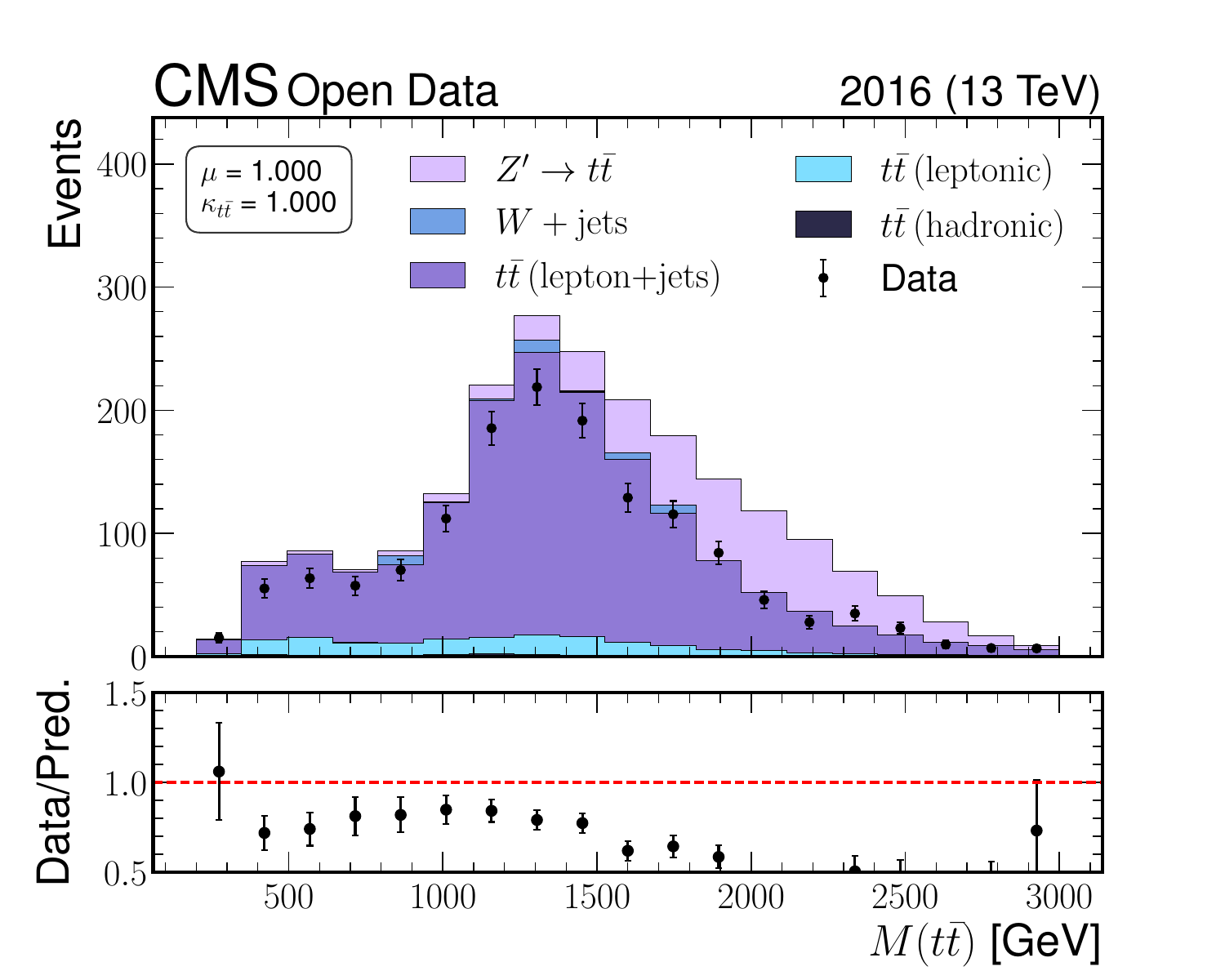}
    \includegraphics[width=0.49\textwidth]{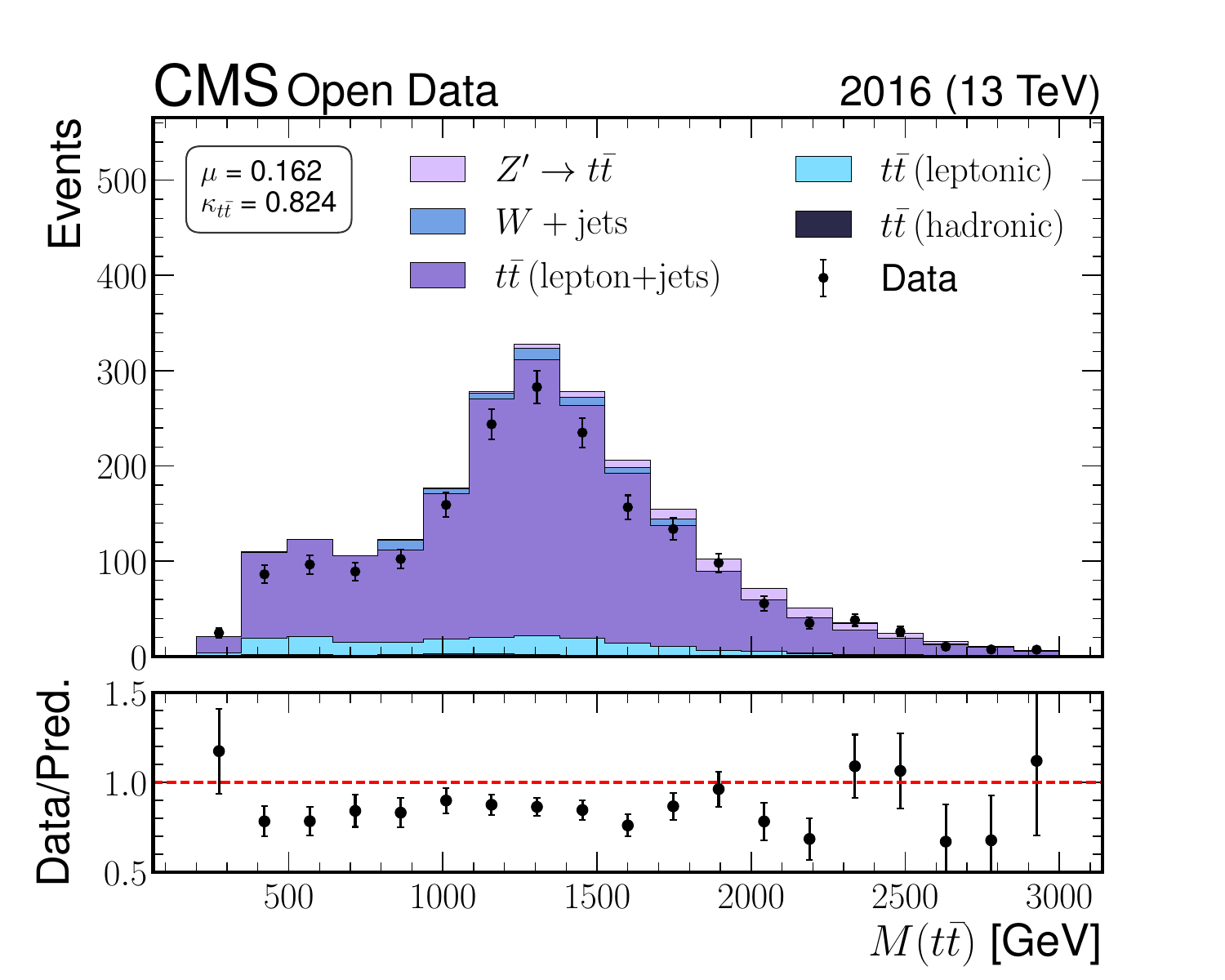}
    \caption{Observed and predicted distributions of the reconstructed invariant mass of the $t\bar{t}$ system after the selection of events as described in Section~\ref{sec:zprime-analysis}. On the left, the prefit distribution before any optimization is shown. The plot on the right shows the postfit distribution after the holistic optimization.}
    \label{fig:mtt}
\end{figure}

\begin{figure}
    \centering
    \includegraphics[width=1.0\textwidth]{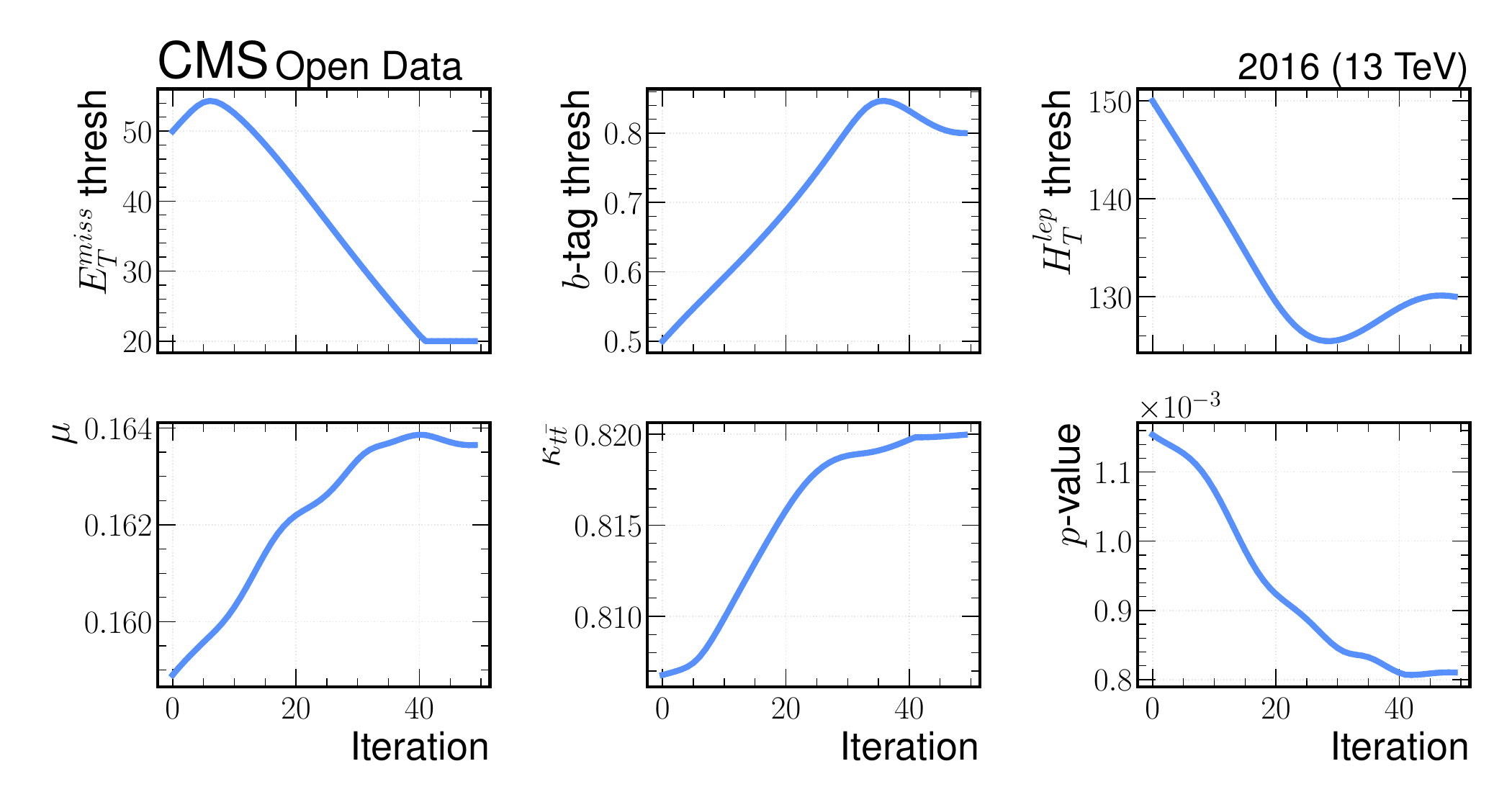}
    \caption{Values of the analysis parameters as a function of the iteration steps of the holistic optimization. The bottom right plot shows the $p$-value, the objective that was being minimized.}
    \label{fig:params}
\end{figure}

\subsection{Challenges, compromises and routes to solutions}

Constructing a fully differentiable pipeline with real data brought forward challenges related to software support and tool integration. For data representation, the  \texttt{JAX} backend of \texttt{awkward-array} is still under active development, and certain operations such as broadcasting and slicing jagged arrays during tracing are not yet supported. This limits the range of observables that can be implemented. A list of missing features is being compiled and shared with the developers to help guide future improvements. In parallel, compatibility issues have been encountered with the \texttt{neos} framework due to environment conflicts with newer dependencies. These constraints currently hinder its integration into the pipeline, although the underlying approach remains valuable.  

Furthermore, the pipeline can in principle handle systematic uncertainties via template histograms. However, implementing the required modifiers in pure \texttt{JAX} proves cumbersome. 
The \texttt{evermore} library, developed in \texttt{JAX} and providing a flexible user interface for likelihoods with systematic uncertainty modifiers, is being explored as a more sustainable solution. Finally, while just-in-time (JIT) compilation in \texttt{JAX} has the potential to provide significant speedups, tracer leaks are observed when applying it to the full workflow. 
These issues are under investigation, and alternative optimizers such as \texttt{optimistix} are being tested to increase stability and efficiency.  

Overall, the challenges encountered highlight both the opportunities and the practical limitations of deploying differentiable pipelines with current tools. This emphasizes the importance of continued collaboration between physicists and software developers.

\section{Optimization results}
The differentiable optimization leads to a substantial improvement in sensitivity with respect to the manual optimization of the analysis design.  
Using the \texttt{OptaxSolver} from \texttt{jaxopt}, the analysis and fit parameters are optimized simultaneously with respect to the discovery $p$-value as described in Section~\ref{sec:implicit_diff}. Figure~\ref{fig:mtt} shows the distribution of the $m_{t\bar{t}}$ histogram before and after the optimization. The good agreement between observation and prediction after optimization demonstrates the validity of this approach.

In 50 iterations, corresponding to roughly two minutes of runtime without JIT compilation, the expected significance improves by roughly 50\% compared to the baseline selection from the CMS open data workshop.  
This demonstrates the effectiveness of gradient-based optimization in navigating the multi-dimensional parameter space of a realistic analysis. Figure~\ref{fig:params} shows the history of the optimized parameters throughout the optimization. 

\begin{figure}[t!]
    \centering
    \includegraphics[width=0.49\textwidth]{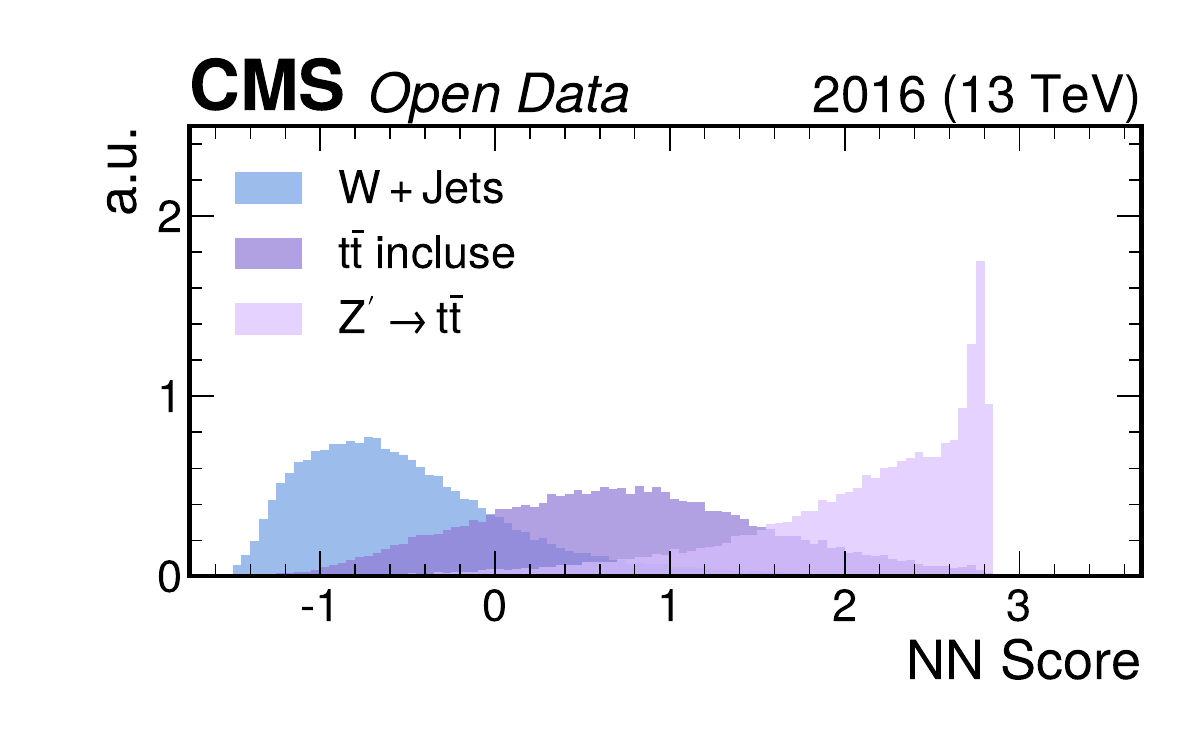}
    \includegraphics[width=0.49\textwidth]{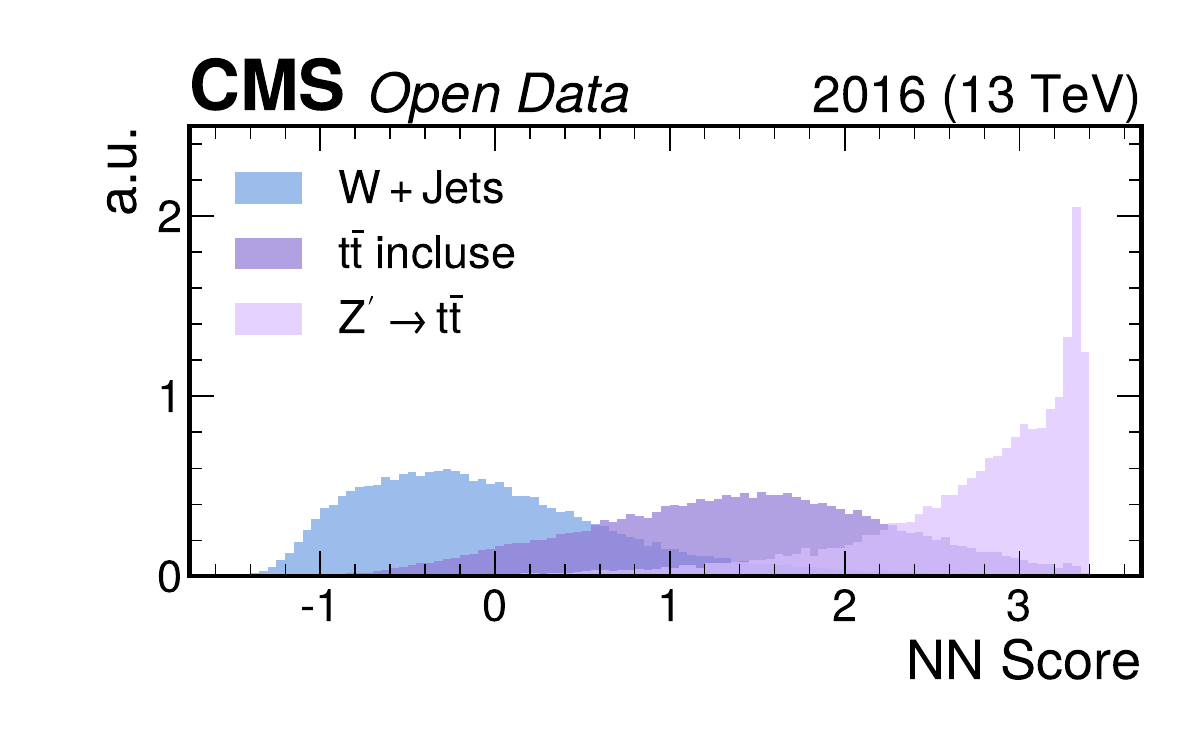}
    \caption{Distribution of the binary classifier's output before (left) and after (right) the holistic optimization.}
    \label{fig:mva_score}
\end{figure}

The framework also allows the holistic optimization of ML components. A simple neural network, initially pre-trained to separate $W$+jets from $t\bar{t}$ events, is fine-tuned as part of the differentiable workflow. Figure~\ref{fig:mva_score} shows the distribution of the classifier output  before and after the optimization. After joint optimization, the classifier shifts towards higher scores in general and in particular, to a slightly better signal-versus-background discriminator. This outcome illustrates the feasibility of optimizing $\mathcal{O}(10^3)$ parameters within a single differentiable pipeline. 

\section{Conclusion and future work}

In this work, the application of differentiable programming to a realistic HEP analysis using CMS open data is demonstrated. 
By constructing an end-to-end differentiable pipeline, the analysis cuts are optimized and result in a substantial improvement in discovery significance with respect to the standard method used in analysis design.
This represents an important step forward in scaling the proof-of-concept introduced in the \texttt{neos} study to realistic physics problems. 
The study also underlines challenges concerning software interoperability and support. Addressing these issues is crucial for the wider adoption of differentiable techniques in HEP analyses.  

Looking ahead, several avenues are being pursued. The analysis will be extended in complexity, incorporating control regions and systematic uncertainties to better reflect a complete HEP workflow. More advanced ML models are going to be integrated and optimized within the pipeline, with the goal of applying these methods to full, publication-quality measurements in collaboration with experimental analyses. 
Meanwhile, continued engagement with the wider community remains essential, through documenting missing features, reporting issues, and contributing improvements that strengthen the differentiable Scikit-HEP ecosystem.  

Overall, differentiable programming offers considerable promise for the future of HEP analyses, providing more powerful and automated ways to extract physics insights from the increasingly large datasets produced at the LHC and future experiments.

\section*{Acknowledgments}
This work is supported by the National Science Foundation under Cooperative Agreements OAC-1836650 and PHY-2323298.

\newpage

\bibliographystyle{JHEP}
\bibliography{bib}

\end{document}